\documentclass[10pt,conference]{IEEEtran}
\IEEEoverridecommandlockouts
\usepackage{amsfonts}

\usepackage{circuitikz}
\usepackage{tikz}
\usepackage{standalone}
\usepackage{tikzscale}
\usetikzlibrary{backgrounds,fit,positioning}
\usetikzlibrary{arrows, patterns,positioning,shapes.geometric,fit,backgrounds}
\usepackage{algorithm} 
\usepackage{algpseudocode} 
\usepackage{mathtools}
\usepackage[thinc]{esdiff}
\usepackage{amsmath,amssymb,color,graphicx,caption,subcaption,comment,tikz,url,latexsym} 
\usepackage{textcomp}
\usepackage{pgfplots}
\usepackage{xcolor}
\usepackage{url}
\usepackage{multirow}
\usepackage{graphicx}
\graphicspath{{.}{Images/}}
\usepackage{caption}
\usepackage{dsfont}
\usepgfplotslibrary{units}
\usetikzlibrary{spy,backgrounds}
\usepackage{pgfplotstable}

\DeclareMathOperator*{\argmin}{argmin}
\DeclareMathOperator*{\argmax}{argmax}
\DeclareMathOperator{\diag}{diag}

\DeclarePairedDelimiter\norm{\lvert\lvert}{\rvert\rvert}

\begin{document}

\title{Adaptive Passive Beamforming in RIS-Aided Communications With Q-Learning
\thanks{This work was supported by the ANR under the France 2030 program, grant NF-PERSEUS : ANR-22-PEFT-0004}
}

\author{\IEEEauthorblockN{Thomas Chêne, Oumaïma Bounhar, Ghaya Rekaya-Ben Othman}
\IEEEauthorblockA{\textit{Communications and Electronics dept.} \\
\textit{Télécom Paris}\\
Palaiseau, France \\
{thomas.chene, oumaima.bounhar, ghaya.rekaya}@telecom-paris.fr}
\and
\IEEEauthorblockN{Oussama Damen}
\IEEEauthorblockA{\textit{Electrical and Computer Engineering dept.} \\
\textit{University of Waterloo}\\
Waterloo, Canada \\
mdamen@uwaterloo.ca}
}

\maketitle

\begin{abstract}
Reconfigurable Intelligent Surfaces (RIS) appear as a promising solution to combat wireless channel fading and interferences. However, the elements of the RIS need to be properly oriented to boost the data transmission rate. In this work, we propose a new strategy to adaptively configure the RIS without Channel State Information (CSI). Our goal is to minimize the number of RIS configurations to be tested to find the optimal one. We formulate the problem as a stochastic shortest path problem, and use Q-Learning to solve it.
\end{abstract}

\begin{IEEEkeywords}
Reconfigurable Intelligent Surfaces (RIS), Bayesian Inference, Q Learning
\end{IEEEkeywords}


\section{Introduction}
\label{Introduction}

With the rapid growth of the Internet of Things (IoT), wireless networks are faced with the challenge of having to handle an unprecedented number of connected devices. A solution lies in the deployment of massive Multiple-Input Multiple-Output (mMIMO) systems. High data rates at millimeter wave (mmWave) frequencies, however suffer from severe signal absorption by obstacles such as buildings \cite{shlezinger_dynamic_2021, wu_towards_2020-1}. Reconfigurable Intelligent Surfaces (RIS) have gained attention for enhancing wireless communication in fading environments \cite{di_renzo_smart_2020}. Built from passive elements, RIS can configure phase shifts to improve signal transmission \cite{pan_reconfigurable_2021-1, liu_reconfigurable_2021}. In the litterature, most work has assumed perfect channel state information (CSI) for optimizing the RIS configuration \cite{wu_intelligent_2019}, although this is very challenging for passive RIS without transceiver chains. Channel estimation is even more complicated with the increase in the number of reflecting elements \cite{bjornson_reconfigurable_2020}. To simplify this, adjacent RIS elements can be grouped to share the same configuration, though this reduces performance \cite{zheng_intelligent_2020_OFDM}. Alternatively, adding a few active elements with receiver chains allows channel estimation but changes the RIS from passive to active \cite{alexandropoulos_hardware_2022}.

Blind methods, such as beam sweeping, exist but require many pilots to be sent. Hierarchical search techniques have been proposed to reduce the number of pilot sent, but they rely on specific codebooks and their performances degrade in low signal-to-noise ratio (SNR) environments. \cite{bjornson_reconfigurable_2020}. 

Recent advances in reinforcement learning (RL) have shown promise in complex optimization tasks \cite{sutton_reinforcement_nodate-1}, and its potential for RIS optimization is emerging \cite{xu_practical_2022-1}. However, many of these approaches still assume knowledge of the channel, a limitation we aim to overcome.

\noindent \emph{Contributions:}
In this work, we introduce an adaptive protocol to maximize the achievable rate for RIS-assisted wireless networks without the need for CSI. We formulate the problem of minimizing the number of pilots sent as a stochastic shortest path (SSP), that to the best of our knowledge, has not been proposed in the literature. We propose to solve the SSP problem using a Q-learning algorithm. We then compare our method to well-known benchmarks.
\\
\noindent \emph{Notation:} We use bold lowercase for vectors, $\mathbf{x}$, bold uppercase letter for matrices $\mathbf{X}$. $(\cdot)^T$ the transpose, $(\cdot)^H$ the hermitian.


\section{Problem-Statement}
\label{Problem-Statement}

\begin{figure}[h]
\centering
    \includegraphics[width=0.4\textwidth]{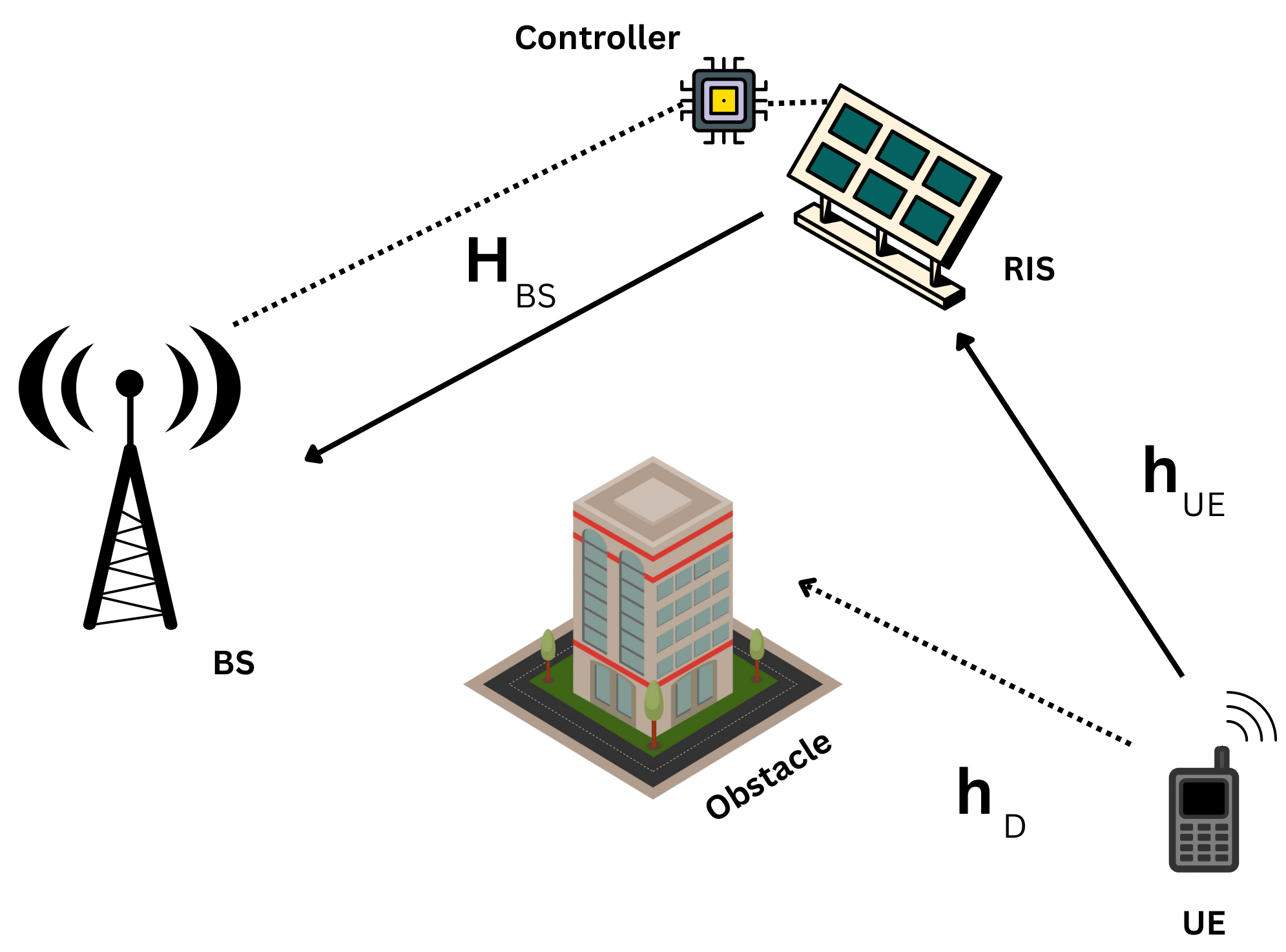}
    \caption{Setup}
    \label{figRIS}
\end{figure}

We are interested in a transmission between a BS with $M$ antennas and a UE with 1 antenna. A RIS with $N$ reflective elements is deployed to facilitate the data transmission. We denote as $\bold{H}_{BS} \in \mathbb C^{N\times M}$, $\bold{h}_{UE} \in \mathbb C^{M\times 1}$ and $\bold{h}_{D} \in \mathbb C^{1\times M}$ the respective channels between the RIS and the BS, between the UE and the RIS and the direct link between the BS and the UE. We further assume that the direct link $\bold{h}_{D}$ is blocked.

\subsection{System Model}

We will consider a mmWave communication system with half-wave spaced uniform linear arrays, as adopted in other previous works \cite{ningTerahertzMultiUserMassive2021}, \cite{xiaoHierarchicalCodebookDesign2016}. The channels can be expressed as:
\begin{equation}
	\begin{split}
            \bold{H}_{BS} &=  \mathbf{a}_{N}(\phi_{1}) \mathbf{a}_{M}^H(\phi_{2}) \\
            \bold{h}_{UE} &=  \mathbf{a}_{1}(\phi_{3}) \mathbf{a}_{N}^H(\phi_{4}) 
    	\end{split}
    	\label{H_RIS_eq}
\end{equation}

where $\mathbf{a}_{N}(\phi)$ is the steering vector function defined as $\mathbf{a}_{N}(\phi) = [1,e^{j\pi \sin (\phi)},..., e^{j\pi (N-1) \sin (\phi)}]^T$, $\phi_{1}$ and $\phi_{2}$ are the angle of arrival (AoA) and angle of departure (AoD) of the channel between the RIS and the BS. $\phi_{3}$ and $\phi_{4}$ are the ones for the channel between the UE and the RIS.

\subsubsection{Received signal}

The signal at the RIS will be linearly transformed by a diagonal matrix $\bold\Theta = \diag(\Phi)$, with $ \Phi =[e^{j\theta_0},....,e^{j\theta_{N-1}}]$, where the $\theta_n$ are the phase shifts introduced by the RIS. Thus the full channel between the BS and the UE is:

\begin{equation}
	\bold h(\Phi) = \bold{H}_{BS} \bold\Theta \mathbf{h}_{UE}
\end{equation}

The signal received at the UE is :
\begin{equation}
	\mathbf y = \sqrt{P}\mathbf h(\Phi) s + \mathbf w
	\label{received}
\end{equation}

where $P$ is the transmit power of the UE, $s\in \mathbb C$ (with $\norm{\mathbf s}^2 = 1$) is the signal sent by the UE, $ \mathbf w \sim \mathcal{CN}(0,\sigma_{w}^2\mathbb I_M)$ an additive white noise.

\subsubsection{Codebook}
We select the RIS reflection matrix from codebooks. We define two types of codebooks:
\begin{itemize}
	\item One codebook $\mathcal C_\Phi = \{\Phi_1,\ldots,\Phi_{N_c}\}$, whose purpose is to maximize the achievable rate between the UE and the BS. Its size is $|\mathcal C_\Phi| = N_c$.
	\item One codebook $\mathcal C_\Psi = \{\Psi_1,\ldots,\Psi_{N_p}\}$, whose purpose is to determine which is the optimal codeword in $\mathcal C_\Phi$, for a unknown channel. Its size is $|\mathcal C_\Psi| = N_p$ 
\end{itemize}
Hence, the reflection matrix of the RIS can take values in the full codebook $\mathcal C_\Phi \cup \mathcal C_\Psi$.

\subsubsection{Rate}
We want to find the codeword in $\mathcal C_\Phi$ that maximizes the achievable rate between the UE and the BS, without CSI:
\begin{IEEEeqnarray}{rCl}
	\max_{\Phi} (\log_2(1+\frac{P||\bold{h}(\Phi)||_2^2}{\sigma_w^2}))
\label{Rate_1_eq}
\end{IEEEeqnarray}
Hence, we will maximize the strength of the channel between the UE and the BS:
\begin{IEEEeqnarray}{rCl}
	\max_{\Phi} ||\bold{h}(\Phi)||_2^2
\label{Channel_strength_1_eq}
\end{IEEEeqnarray}


\section{Proposed-Approach}
\label{Proposed-Approach}

\subsection{Protocol}
We assume that no prior knowledge of the channels $\bold{H}_{BS}$ and $\bold{h}_{UE}$ is accessible. Hence, we cannot solve an optimization problem to find $\Phi$ that maximizes the rate. 
We need to test configurations $\Psi \in \mathcal C_\Psi$ in order to increase our knowledge of which $\Phi \in \mathcal C_\Phi $ maximizes the rate. The BS will compute an indicator of the quality of the channel between the UE and BS. We assume that the BS computes the received signal energy:

\begin{equation}
	Y = \mathbf y \mathbf y^{H}
\end{equation}

We model this feedback received at the BS as :
\begin{equation}
	Y = f(\Phi,\bold H,\mathbf w) \in \mathbb{R}
	\label{modeled_received}
\end{equation}
For sake of simplicity we wrote $\mathbf H = \{\bold{H}_{BS},\bold{h}_{UE}\}$ to represent the overall unknown channel. With $\Phi \in \mathcal C_\Phi \cup \mathcal C_\Psi$, $\mathbf w$ the same noise as in \eqref{received}. 

Hence, the received signal is a noisy function of unknown parameters $\mathbf H$ (the channel), and known parameters $\Phi$ (the configuration of the RIS). The higher the value of $Y$, the ``better'' the configuration of the RIS is.

\subsubsection{Searching for the optimal codeword}
 At every time instant $k$, the UE will send a symbol $\mathbf s$, the RIS uses codeword $\Psi_{A_k}$ ($A_k$ is the index of the codeword to use). The BS will receive a signal $\mathbf y_k$ and computes $Y_k$. We schematize it in Fig. \ref{figRIS}. After $L_p$ iterations, we used codewords $\mathbf{\Psi}_{A_1}^{A_{L_p}} = [\Psi_{A_1},\ldots,\Psi_{A_{L_p}}]$, and received samples at the BS $\mathbf Y_{1}^{L_p} = [Y_1,\ldots,Y_{L_p}]$.

\begin{figure}[h]
	\centering
         \resizebox{\columnwidth}{!}{%
          \includegraphics[width=7cm]{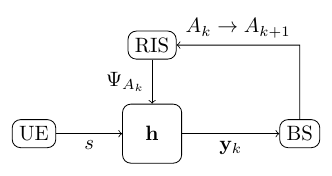}%
        }
		
	\caption{Adaptive Protocol}
	\label{figRIS}
\end{figure}

\subsubsection{Adaptive search}
We define an ``acquisition" function $\mathcal A(\cdot)$, that takes into account the previously received feedbacks $\mathbf Y_{1}^{k-1} = [Y_1,\ldots,Y_{k-1}]$ and codewords tested $\mathbf{\Psi}_{A_1}^{A_{k-1}} = [\Psi_{A_1},\ldots,\Psi_{A_{k-1}}]$, and that will give the index of the next codeword to use $A_k$:
\begin{equation}
	\mathcal A_k\colon \{\mathbb R\}^{k-1} \otimes \{[1;N_p]\}^{k-1} \rightarrow [1;N_p]
\end{equation}  

Hence, we receive:
\begin{IEEEeqnarray}{rCl}
	Y_{1} &&= f(\Psi_{\mathcal A_1},\mathbf H,\mathbf w_1) \nonumber\\
    Y_{2} &&= f(\Psi_{\mathcal A_2(Y_1,\Psi_{A_1})},\mathbf H,\mathbf w_2) \nonumber\\
    \cdots \nonumber\\
    Y_{k} &&= f(\Psi_{\mathcal A_k(\mathbf Y_1^{k-1},\mathbf{\Psi}_{A_1}^{A_{k-1}})},\mathbf H,\mathbf w_k)\nonumber
\end{IEEEeqnarray}
In Fig. \ref{figRIS}, after the BS receives $\mathbf y_k$, the configuration of the RIS will be updated from $\Psi_{A_k}$ to $\Psi_{A_{k+1}}$.
\subsection{Classification}
\label{Classification}

\subsubsection{Optimal codeword}
We define the class of a channel:
\begin{IEEEeqnarray}{rCl}
	\Phi(\mathbf H) = \{\Phi \in \mathcal C_\Phi \mid \Phi = \argmax_{\Phi \in \mathbf\Phi}(||\bold{h}(\Phi)||_2^2)\}.
\end{IEEEeqnarray}
This corresponds to the best codeword of a codebook written as a function of $\mathbf H$.

\subsubsection{Correct classification}
After having received $L_p$ feedbacks, we will declare a codeword $\Phi_{dec}$ that we consider to be the best codeword. We compare the resulting strength of the channel that we obtain when using $\Phi_{dec}$, to the one when using $\Phi(\mathbf H)$:

\begin{equation}
    ||\bold{h}(\Phi_{dec})||_2^2 > p(\Phi_{dec}=\Phi(\mathbf H)) \max_{\Phi \in \mathbf\Phi}(||\bold{h}(\Phi)||_2^2)
 \label{correct_channel}
\end{equation}

By fixing a probability of correct classification higher than a fixed degree of precision $1-\delta$, $\delta \in [0;1]$:

\begin{equation}
	p_{correct}(\mathcal A,L_p) = p(\Phi_{dec}=\Phi(\mathbf H)) > 1-\delta
 \label{correct_higher}
\end{equation}

We obtain:

\begin{equation}
    \frac{||\bold{h}(\Phi_{dec})||_2^2}{\max_{\Phi \in \mathbf\Phi}(||\bold{h}(\Phi)||_2^2)}>(1-\delta)
 \label{correct_channel_2}
\end{equation}

By correctly declaring the optimal codeword, we can maximize the channel strength.
\subsubsection{Bayes Optimal Classifier}
We define:
\begin{IEEEeqnarray}{rCl}
    && p(\Phi(\mathbf H)| \mathbf Y_{1}^{L_p},\mathbf{\Psi}_{A_1}^{A_{L_p}}) \nonumber\\
    && := [p(\Phi_1| \mathbf Y_{1}^{L_p},\mathbf{\Psi}_{A_1}^{A_{L_p}}), \cdots, p(\Phi_{N_c}| \mathbf Y_{1}^{L_p},\mathbf{\Psi}_{A_1}^{A_{L_p}})]
\end{IEEEeqnarray}
Given the received samples, the Bayes optimal classifier is:

\begin{equation}
	\argmax_{c \in [1;N_c]} p(\Phi(\mathbf H)| \mathbf Y_{1}^{L_p},\mathbf{\Psi}_{A_1}^{A_{L_p}})
\end{equation}

Using this classifier, the probability of correct classification when using the acquisition function $\mathcal A$, and having received $L_p$ feedbacks is:
\begin{equation}
	p_{correct}(\mathcal A,L_p) = \max_{c \in [1;N_c]} p(\Phi(\mathbf H)| \mathbf Y_{1}^{L_p},\mathbf{\Psi}_{A_1}^{A_{L_p}})
\end{equation}

Hence, as in \eqref{correct_higher} we want:
\begin{equation}
	\max_{c \in [1;N_c]} \mathbf p(\Phi(\mathbf H)| \mathbf Y_{1}^{L_p},\mathbf{\psi}_{A_1}^{A_{L_p}}) > 1-\delta
 \label{cond_proba}
\end{equation}
By using this classifier, we are guaranteed to obtain a channel strength proportional to the optimal one, up to a factor $1-\delta$ as in \eqref{correct_channel_2}.
\subsubsection{Minimizing the number of pilots}
With more feedbacks we could on average increase even further the probability of correct classification (information cannot hurt), but we only want to guarantee the classification up to a degree of precision $\delta$ :
\begin{equation}
	L_{min}(\mathcal A,\mathbf H,\mathbf w) = \min \{L_p / p_{correct}(\mathcal A,L_p)>1-\delta\}
\end{equation}

We need to find a function $\mathcal A$ that will guarantee correct classification up to a certain degree of precision $\delta$, with few feedbacks, for all possible $\mathbf H$:
\begin{equation}
	L^{*} = \min_{\mathcal A} \mathbb E_{\mathbf H,\mathbf w}(L_{min}(\mathcal A,\mathbf H,w))
 \label{ssp_pb}
\end{equation}

\subsection{Stochastic Shortest Path problem}

\begin{figure}[h]
    \centering
    \includegraphics[width=6cm]{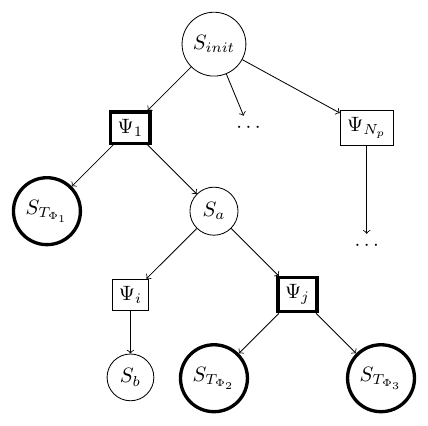}
    \caption{Stochastic shortest path problem}
    \label{Graph}
\end{figure}

We represent graphically the problem in fig.\ref{Graph}. When no samples are received and no knowledge of the channel is available, we start in a state that we call $S_{init}$. Testing a codeword $\Psi_k$ allows us to receive a new sample and move in the graph. The state that we will reach depends on the action $\Psi_k$ and the unknown channel $\mathbf H$. In Fig.\ref{Graph} for example, taking the action $\Psi_1$ can bring us into different states. When we reach a state that satisfies \eqref{cond_proba} (we call those terminal states $\mathcal S_{T_\Phi}$) we can stop sending feedbacks. The length of the path between the initial state and the terminal state for a channel $\mathbf H$ when using acquisition function $\mathcal A$ is $L_{min}(\mathcal A,\mathbf H,\mathbf w)$, the best average length of the path is $L^\star$. The problem we are trying to solve is a \textbf{Stochastic Shortest Path} problem (SSP).

\subsection{Markov Decision Process}
\label{MDP}
To solve the SSP problem we formulate it as a Markov Decision Process (MDP). An MDP is defined by its State space, Action Space, Reward, and Transition probability \cite{sutton_reinforcement_nodate-1}:

\begin{itemize}
    \item \textbf{State Space} : $S$, denotes a state from the state space. Each state corresponds to a probability (a vector of size $N_c$), $S = p(\Phi(\mathbf H)| \mathbf Y,\mathbf{\Psi})$ for some received feedbacks $\mathbf Y$ and some codewords tested $\mathbf{\Psi}$. The state $S$ represents the knowledge that we have acquired by testing different configurations of the RIS. The agent starts at the initial state $S_{init} = p(\Phi(\mathbf H)) = [\frac{1}{N_c},\cdots,\frac{1}{N_c}] $ (for example we assumed that without CSI, the probability is uniformly distributed among the codewords). The goal is to reach a terminal state where the probability respects \eqref{cond_proba}. We call those states, $\mathcal{S}_{T_{\Phi_i}} = \{ \mathcal S / p(\Phi_i| \mathbf Y,\mathbf{\Psi}) > 1 - \delta\}$. \\

    \item \textbf{Action Space} : the action space is defined by $[1,N_p]$ (the indices of the codebook $\mathcal C_\Psi$). Each action corresponds to testing a codeword $\Psi$. \\

    \item \textbf{Reward} : $R_{A}(S, S') \in  \{0,-1\}$, the received reward  after transitioning from state \(S\) to state \(S'\) due to action \(A\). The reward is $0$ when we reach a terminal state and $-1$ otherwise. \\

    \item \textbf{Transition probability} : $\mathbf{P}_{A}(S,S') = P(S_{t+1} = S' \mid S_{t} = S, A_{t} = A))$ is the probability of being in the state \(S'\) at time \(t+1\) after taking the action \(A\) from the state \(S\) at time \(t\).
\end{itemize}

\subsection{Optimal Policy}
Because the reward received by the agent at time $t$ is $-1$ as long as the terminal state is not reached, the sum of the rewards corresponds to the opposite of the length of the path between the initial state and the terminal state:
\begin{equation}
    L_{min}(\pi,\mathbf H,\mathbf w) = -\sum_{t=0}^{\infty} R_t
\end{equation}
with $R_t$ the rewards received when starting from the state $S_{init}$, $\pi$ corresponds to the ``policy" we want to learn, that we previously called ``$\mathcal A$".
Hence we rewrite our objective \eqref{ssp_pb} with the MDP formalism:
\begin{equation}
    L^{*} = -\max_{\pi} \mathbb{E}_{\mathbf H,\mathbf w}\left[\sum_{t=0}^{\infty} R_t \right]
\end{equation}


\section{Q-Learning}
\label{Preliminaries_Reinforcement_learning}

\begin{figure}[t]
\centering
    \includegraphics[width=8cm]{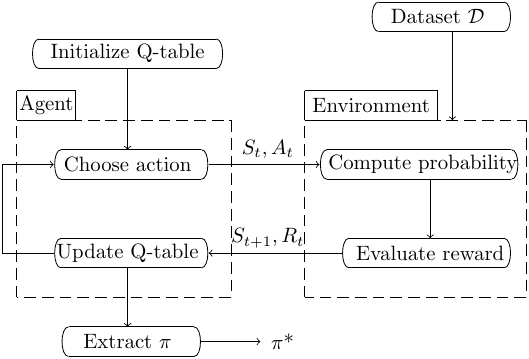}
    \caption{Training Procedure of Q-Learning}
    \label{fig:Q}
\end{figure}

To find the optimal policy, we train an agent, as schematized in fig. \ref{fig:Q}:
\begin{itemize}
    \item The agent receives an observation $S_t$ at each time step $t$,
    \item The agent chooses an action $A_t$ according to the observation,
    \item The environment  transition to a new state $S_{t+1}$,
    \item The agent obtains a reward ${R}_{A_t}(S_t, S_{t+1})$.
\end{itemize}

\subsection{Bellman Equation}
The Q-Learning algorithm uses the state-action value function, also called the \textbf{Q-function}, and defined as : 

\[
Q(S_t, A_t) = \mathbb{E}_{\mathbf H,\mathbf w}\left[\sum_{k=0}^{\infty} R_{t+k+1} \mid S_t = S, A_t = A\right]
\]

To maximize the Q-function, the Q-Learning algorithm is based on the use of the Bellman Equation (\ref{eqn:bellman}). It is a recursive equation, that is iteratively updated during the training phase. 

\begin{IEEEeqnarray}{rCl}
Q(S_t, A_t) \xleftarrow\ &&(1-\alpha) Q(S_t, A_t) \nonumber \\
&&+ \alpha[{R}_{t+1} + max_{A}Q(S_{t+1}, A_t)]
\label{eqn:bellman}
\end{IEEEeqnarray}
with $\alpha \in [0,1]$, the learning rate.

\subsection{Dataset}
We assume that we have access to a dataset to train the Q-Learning, that can be obtained by storing feedbacks from the UE for different channels, we call the dataset $\mathcal{D}$ = $(\mathbf X,\mathbf S_{T_{Real}})$: 

\begin{itemize}
    \item The received feedbacks: \\
    \raggedright $\mathbf X = [\mathbf X(\mathbf H_1),\ldots,\mathbf X(\mathbf H_{N_{Dataset}})]$, where $\mathbf X(\mathbf H_k) = [f(\Psi_1,\bold H_k,\mathbf w_{1,k}),\ldots,f(\Psi_{N_p},\bold H_k,\mathbf w_{N_p,k})]$ for $k \in [1,N_{Dataset}]$.
    
    \item The real terminal states: \\
    \raggedright $\mathbf S_{T_{Real}} = [\mathbf S_{T}(\mathbf H_1),\ldots,\mathbf S_{T}(\mathbf H_{N_{Dataset}})]$, where $\mathbf{S}_T(\mathbf H_k) = [0 \cdots 0 \underset{\substack{\downarrow \\ \text{\it $c^{*}$}}}{1} 0  \cdots 0 ] $,
    $c^{*} = \argmax_{c \in [1;N_c]} ||\bold{h}(\Phi_c)||_2^2)$.
\end{itemize}

\subsection{Implementation of the Q-Learning algorithm}
The algorithm \ref{Alg1} describes the implementation.

\begin{algorithm} 
	\begin{algorithmic}[2]
        \item \textbf{Inputs:} Training dataset $\mathcal{D}$, State space $\mathbf{S}$, Terminal states $\mathbf{S}_{T_{\Phi}} \in \mathbf{S}$, Action space $\mathbf{A}$. Hyperparameters : $\epsilon$, $\alpha$
        \item \textbf{Output:} Policy $\pi$
        \item Initialize: Q-matrix
        \State \underline{Step 1:} Learn the Q-matrix
        \For{$epoch = 1$ to $max\_epoch$} 
            \For{$h = 1$ to $max\_channel$}
                \State Choose an element in the dataset $\mathbf X(\mathbf H_k)$
                \State Start at $S_{init}$, $Y_{tot}$ = \{$\emptyset$\}
                \For{$L = 1$ to $max\_L$}
                    \State \underline{Step 1:} Epsilon Greedy: $r \sim \mathcal U[0,1]$
                    \If{$r > \epsilon$}
                        \State $A \gets \arg\max_{A} Q(S, A)$
                    \Else
                        \State $A \gets \text{random choice from Action space $\mathbf{A}$}$
                    \EndIf
                    \State $Y_{l}$ = $X(\mathbf H_k)[A]$
                    \State $Y_{tot} \gets Y_{tot} \cup \{Y_{l};A\}$
                    \State $\mathbf{p} \gets p(\Phi \mid Y_{tot})$
                    \State Next state $S \gets \argmin_{S \in \mathbf{S}} ||\mathbf{p}-S||_2^2$
                    \State $Q(S,A) \gets $ Update with the Bellman eq. (\ref{eqn:bellman})
                    \If{$S \in \mathbf{S_{T}}$}
                            \State reward = 0
                            \State break
                        \Else
                            \State reward = -1
                        \EndIf
                \EndFor
            \EndFor
        \EndFor
        \State \underline{Step 2:} Extract the policy
        \For{$s = 1$ to $| \mathbf S |$}
            \If{$\argmax_{A} Q(s,A) = a$}
                \begin{equation}
                    \pi_{opt}(s,a) \gets 1 
                \end{equation}
            \EndIf
        \EndFor
	\end{algorithmic} 
\caption{Q-Learning algorithm}
\label{Alg1}
\end{algorithm}

\section{Probabilities and finite state size}
\label{Computing_probabilities}

\subsection{Approximation of the probability}
We use the euclidean distance between the samples $\mathbf Y_{1}^{L_p}$ we received and the elements of our dataset $\mathcal D$ to compute the probability:

\begin{IEEEeqnarray}{rCl}
    &&p(\Phi(\mathbf H) = \Phi_k| \mathbf Y_{1}^{L_p},\mathbf{\Psi}_{A_1}^{A_{L_p}}) \propto \nonumber\\
    &&p(\Phi(\mathbf H) = \Phi_k| \mathbf Y_{1}^{L_p-1},\mathbf{\Psi}_{A_1}^{A_{L_p-1}}) \nonumber \\
    &&\cdot\sum_{\mathbf H_l / \mathbf S_{T}(\mathbf H_l)[k] = 1}  
    \frac{1}{(\mathbf X(H_l)[A_{L_p}]-\mathbf Y_{L_p})^2}
    \label{approx_proba}
\end{IEEEeqnarray}
With $\propto$ meaning proportional, we will normalize every component of the probability so that it sums up to one.

\subsection{Finite state size}
When using Q-Learning, the number of states needs to be finite, hence we need to discretize the state-space. The initial state and terminal states are described in \ref{MDP}. Other states are assumed to approximate the most likely probability distributions. The more we discretize the space, the worse the Q-Learning algorithm will perform. The choice of discretization of the space has consequences on the performances but is not the primary focus of the paper. We propose to discretize the space using states such as ($[0,\cdots,0,1-q,0\cdots,0,q,\cdots,0)]$, for different values of $q \in \{0.1,\ldots,0.9\}$. We use this quantization to represent the fact that at each instant $k$, we have two states that are likely to be the optimal ones, and we will make an action to resolve the uncertainty.


\section{Numerical Results}
\label{Numerical Results}

\subsection{Numerical setup}
We numerically evaluate the proposed method. We set the number of antennas at the BS to M = 64, the UE has 1 antenna and the RIS is composed of N = 100 reflective elements. The channel model is the same as in (\ref{H_RIS_eq}). With $\phi_i \sim \mathcal U[0,2\pi]$,$\forall i$. For the different simulations, we generate 1000 different realisations of the channel. We want to compare the performance of our algorithm with a hierarchical method, hence we use a codebook $\mathcal C_\Psi$ that is hierarchical. We use the hierarchical codebook defined for classic beamforming in \cite{xiaoHierarchicalCodebookDesign2016} called DEACT, it was also used for passive beamforming with RIS in \cite{ningTerahertzMultiUserMassive2021} and called Phase Shift Deactivation (PSD), the codebook $\mathcal C_\Psi$ is a binary hierarchical codebook composed of 14 beams. The codebook $\mathcal C_\Phi$ is composed of the 8 narrow beams from the hierarchical codebook.

\subsection{Algorithm}
We describe in Algorithm \ref{alg1} the different steps of our method that we described in previous sections.
\begin{algorithm} 
	\caption{Adaptative Blind Beamforming algorithm} 
    \label{alg1}
	\begin{algorithmic}[1]
        \item \textbf{Inputs:} Codebooks $\mathcal C_\Phi$,$\mathcal C_\Psi$, Acquisition function $\mathcal A$ learned with Q-Learning, \textit{algorithm-type}: ``Random Acquisition" or ``Q-Learning Acquisition"
        \item \textbf{Output:} Codeword $\Phi_{L_p} \in \mathcal C_\Phi$ that maximizes the rate
        \item $\mathbf Y = \{\emptyset\}$
        \item $\mathbf \Psi = \{\emptyset\}$
        \For{$k = 1, ..., L_p$} 
            \State \underline{Step 1:} Determine the codeword $\Psi_{k}$:
            \If{\textit{algorithm-type}: ``Random Acquisition"}
                \State $\Psi_{k} = \Psi_{\mathbf\Pi(k)} $
            \ElsIf{\textit{algorithm-type}: ``Q-Learning Acquisition"}
                \State $\Psi_{k} = \Psi_{\mathcal A_k(\mathbf Y,\mathbf \Psi)} $
            \EndIf
            \State \underline{Step 2:} Receive feedback with codeword: $\Psi_{k}$
            \State Receive: $ Y_k = f(\Psi_{k},\mathbf H,\mathbf w_k)$
            
            \State \underline{Step 3:} Update $\mathbf Y$ and $\mathbf \Psi$ by appending $\{Y_k,\Psi_k\}$ to it and update the probability $p(\Phi(\mathbf H) | \mathbf Y,\mathbf{\Psi})$ with \eqref{approx_proba}
        \EndFor
        \item return $\Phi_{\argmax p(\Phi(\mathbf H) | \mathbf Y,\mathbf{\Psi})}$
	\end{algorithmic} 
\end{algorithm}

\subsection{Shortest path $L^{*}$}
\begin{figure}[h]
    \includegraphics[width=8cm]{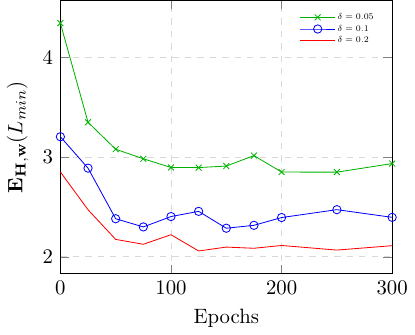}
    \caption{Evolution of the average length $L_{min}$ during the different epochs of the Q-Learning}
    \label{NumberpilotsL_star}
\end{figure}

We want to find an acquisition function $\mathcal A$ that will minimize the average length of the path between a state where we have no knowledge of the optimal codeword, to a state where we have a high probability of finding the optimal codeword. This function is learned through Q-Learning. We plot the average length of the path with the number of epochs. We compare the plot for different values of $\delta$ in Fig. \ref{NumberpilotsL_star}.

We notice in Fig. \ref{NumberpilotsL_star}, that for a smaller $\delta$, the average length increases. Indeed, in order to reach a higher certainty about which codeword is the best, more feedbacks needs to be received. Also, we notice that the difference between the average length at the beginning of Q-Learning (which is a random policy), and the end (for a improved policy) is higher for smaller $\delta$.

\subsection{Comparison with benchmarks}
\begin{itemize}
    \item Exhaustive Search: We use all 8 codewords in $\mathcal C_\Phi$ and then take the best
    \item Hierarchical Search: The hierarchical search is a dichotomic serch that will use $2\log_2(8) = 6$ codewords.
    \item Random Acquisition: $\mathcal A_k$ is random
    \item Q-Learning Acquisition: $\mathcal A_k$ is found with Q-Learning
\end{itemize}

\begin{figure}[h]
    \includegraphics[width=8cm]{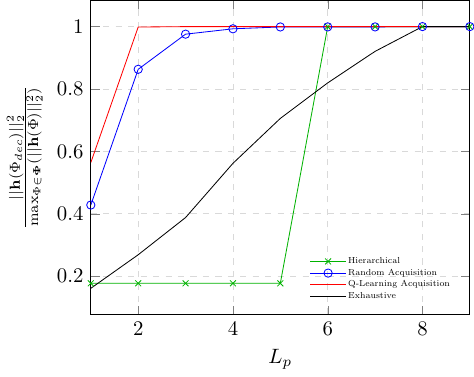}
    \caption{Number of pilots required to reach a given channel strength, for different methods, SNR = 20 dB}
    \label{NumberpilotsL_p}
\end{figure}

As in \eqref{correct_channel_2} we look at the strength of the channel that we obtain by using $L_p$ configurations in Fig. \ref{NumberpilotsL_p} , using our method ($\delta = 0.9$), and compared to the other benchmarks described. We notice that by using a probabilistic formalism, we can reach a high channel strength with fewer configurations tested compared to other methods. Using Q-Learning allows to only use the codewords suited to improve our knowledge of the most likely codeword. Randomly testing codewords proves to be less efficient since it takes more time to converge towards the correct class.

The reason why the plot of the Q-Learning method goes higher than $0.9$ is due to two reasons. First, equation \eqref{correct_channel_2} is an inequality, even though we do not find the optimal codeword, the codeword declared has a non-zero channel strength. Second, the approximation of the probability is not exact, which means that even though the maximum of our approximation is smaller than $1-\delta$, the maximum of the ``real'' probability might be higher.

\section{Conclusion}
In this paper, we proposed a probabilistic method combined with the Q-learning algorithm to optimize the configuration of RIS without requiring CSI. Our approach outperforms traditional blind  methods from the literature, in terms of number of configuration tested. However, the state space grows in the codebook size, and quantifying it accurately becomes increasingly challenging. To address this limitation, future work will explore deep learning techniques to bypass the need for manual state space design, further enhancing the scalability and efficiency of our solution.

\bibliographystyle{ieeetr}
\bibliography{Biblio.bib}

\begin{thebibliography}{10}

\bibitem{shlezinger_dynamic_2021}
N.~Shlezinger, G.~C. Alexandropoulos, M.~F. Imani, Y.~C. Eldar, and D.~R.
  Smith, ``Dynamic {Metasurface} {Antennas} for {6G} {Extreme} {Massive} {MIMO}
  {Communications},'' {\em IEEE Wireless Communications}, vol.~28,
  pp.~106--113, Apr. 2021.

\bibitem{wu_towards_2020-1}
Q.~Wu and R.~Zhang, ``Towards {Smart} and {Reconfigurable} {Environment}:
  {Intelligent} {Reflecting} {Surface} {Aided} {Wireless} {Network},'' {\em
  IEEE Communications Magazine}, vol.~58, pp.~106--112, Jan. 2020.

\bibitem{di_renzo_smart_2020}
M.~Di~Renzo, A.~Zappone, M.~Debbah, M.-S. Alouini, C.~Yuen, J.~De~Rosny, and
  S.~Tretyakov, ``Smart {Radio} {Environments} {Empowered} by {Reconfigurable}
  {Intelligent} {Surfaces}: {How} {It} {Works}, {State} of {Research}, and
  {The} {Road} {Ahead},'' {\em IEEE Journal on Selected Areas in
  Communications}, vol.~38, pp.~2450--2525, Nov. 2020.

\bibitem{pan_reconfigurable_2021-1}
C.~Pan, H.~Ren, K.~Wang, J.~F. Kolb, M.~Elkashlan, M.~Chen, M.~Di~Renzo,
  Y.~Hao, J.~Wang, A.~L. Swindlehurst, X.~You, and L.~Hanzo, ``Reconfigurable
  {Intelligent} {Surfaces} for {6G} {Systems}: {Principles}, {Applications},
  and {Research} {Directions},'' {\em IEEE Communications Magazine}, vol.~59,
  pp.~14--20, June 2021.

\bibitem{liu_reconfigurable_2021}
Y.~Liu, X.~Liu, X.~Mu, T.~Hou, J.~Xu, M.~Di~Renzo, and N.~Al-Dhahir,
  ``Reconfigurable {Intelligent} {Surfaces}: {Principles} and
  {Opportunities},'' {\em IEEE Communications Surveys \& Tutorials}, vol.~23,
  no.~3, pp.~1546--1577, 2021.

\bibitem{wu_intelligent_2019}
Q.~Wu and R.~Zhang, ``Intelligent {Reflecting} {Surface} {Enhanced} {Wireless}
  {Network} via {Joint} {Active} and {Passive} {Beamforming},'' {\em IEEE
  Transactions on Wireless Communications}, vol.~18, pp.~5394--5409, Nov. 2019.

\bibitem{bjornson_reconfigurable_2020}
E.~Bj{\"o}rnson, {\"O}.~{\"O}zdogan, and E.~G. Larsson, ``Reconfigurable
  {Intelligent} {Surfaces}: {Three} {Myths} and {Two} {Critical} {Questions},''
  {\em IEEE Communications Magazine}, vol.~58, pp.~90--96, Dec. 2020.
\newblock arXiv:2006.03377 [cs, eess, math].

\bibitem{zheng_intelligent_2020_OFDM}
B.~Zheng and R.~Zhang, ``Intelligent {Reflecting} {Surface}-{Enhanced} {OFDM}:
  {Channel} {Estimation} and {Reflection} {Optimization},'' {\em IEEE Wireless
  Communications Letters}, vol.~9, pp.~518--522, Apr. 2020.

\bibitem{alexandropoulos_hardware_2022}
G.~C. Alexandropoulos and E.~Vlachos, ``A {Hardware} {Architecture} for
  {Reconfigurable} {Intelligent} {Surfaces} with {Minimal} {Active} {Elements}
  for {Explicit} {Channel} {Estimation},'' May 2022.
\newblock arXiv:2002.10371 [cs, eess, math].

\bibitem{sutton_reinforcement_nodate-1}
R.~S. Sutton and A.~G. Barto, ``Reinforcement {Learning}: {An}
  {Introduction},''

\bibitem{xu_practical_2022-1}
W.~Xu, J.~An, L.~Gan, and H.~Liao, ``A {Practical} {Design} {Based} on {Deep}
  {Reinforcement} {Learning} for {RIS}-{Assisted} {mmWave} {MIMO} {Systems},''
  in {\em 2022 {IEEE} 8th {International} {Conference} on {Computer} and
  {Communications} ({ICCC})}, (Chengdu, China), pp.~1599--1602, IEEE, Dec.
  2022.

\bibitem{ningTerahertzMultiUserMassive2021}
B.~Ning, Z.~Chen, W.~Chen, Y.~Du, and J.~Fang, ``Terahertz {{Multi-User Massive
  MIMO With Intelligent Reflecting Surface}}: {{Beam Training}} and {{Hybrid
  Beamforming}},'' {\em IEEE Transactions on Vehicular Technology}, vol.~70,
  pp.~1376--1393, Feb. 2021.

\bibitem{xiaoHierarchicalCodebookDesign2016}
Z.~Xiao, T.~He, P.~Xia, and X.-G. Xia, ``Hierarchical {{Codebook Design}} for
  {{Beamforming Training}} in {{Millimeter-Wave Communication}},'' {\em IEEE
  Transactions on Wireless Communications}, vol.~15, pp.~3380--3392, May 2016.

\end{thebibliography}

\end{document}